\documentclass[a4paper]{article}

\usepackage{graphicx}

\usepackage{hyperref}
\usepackage{multirow}
\usepackage{subcaption}
\usepackage{amssymb}
\usepackage{algorithm} 
\usepackage{algpseudocode} %
\begin{document}

\title{Introducing Latent Timbre Synthesis}


\author{K{\i}van\c{c} Tatar  \\   Simon Fraser University  \\ Vancouver, B.C., Canada \\ \href{mailto:ktatar@sfu.ca}{ktatar@sfu.ca} \and Daniel Bisig  \\ Zurich University of the Arts \\ Zurich, Switzerland \\ \href{mailto:daniel.bisig@zhdk.ch}{daniel.bisig@zhdk.ch}   \and Philippe Pasquier \\ Simon Fraser University \\ Vancouver, B.C., Canada \\ \href{mailto:pasquier@sfu.ca}{pasquier@sfu.ca}
}

\maketitle

\begin{abstract}

We present the Latent Timbre Synthesis (LTS), a new audio synthesis method using Deep Learning. The synthesis method allows composers and sound designers to interpolate and extrapolate between the timbre of multiple sounds using the latent space of audio frames. We provide the details of two Variational Autoencoder architectures for LTS, and compare their advantages and drawbacks. The implementation includes a fully working application with graphical user interface, called \textit{interpolate\_two}, which enables practitioners to explore the timbre between two audio excerpts of their selection using interpolation and extrapolation in the latent space of audio frames. Our implementation is open-source, and we aim to improve the accessibility of this technology by providing a guide for users with any technical background. 
\end{abstract}

\section{Introduction}
\label{intro}

Promising new research of Deep Learning (DL) for the musical applications of audio transformation and sound synthesis has recently emerged in academia; in conjunction with the increasing popularity of Deep Learning architectures \cite{briot_deep_2020}. Musical applications of these technologies have yet to become accessible for composers and musicians who lack expertise in Machine Learning (ML) and Artificial Intelligence (AI). Although Deep Learning has been applied to many musical tasks \cite{briot_deep_2020}, the research on incorporating Deep Learning architectures for sound design applications in experimental electronic music is still in its early stages. 


This project focuses on the integration of modern ML and AI techniques into tools for computer-assisted sound design and their applications within composition practices. We concentrate on the fields of experimental electronic music and sound art, which takes sound qualities, listening modalities and experiences \cite{tuuri_formulating_2012}, and organized sound theories \cite{roads_composing_2015} into the center \cite{tatar_musical_2019}. We specifically focus on audio corpus-based sound synthesis approaches that rely on large libraries of audio excerpts.  

Latent Timbre Synthesis (LTS) aims to help composers by utilizing an abstract latent timbre space that is generated by training unsupervised Deep Learning (DL) models with a set of audio recordings. These new DL tools allow composers to synthesize sounds using a latent space of audio that is constrained to the timbre space of the audio recordings in the training set. Our development of ML and AI tools for computer-assisted sound design and their subsequent evaluation within composition practice serve to highlight the benefits and shortcomings of the selected machine learning algorithms for creative ideation and discovery. 




In 1940s, physicist Dennis Gabor proposed \cite{gabor_acoustical_1947} that a sound is composed of acoustical quanta that is bounded by time and frequency. We are inspired by this idea while applying this approach to digital audio, and asking,

\begin{itemize}
    \item What would be the latent space of audio frames?
    \item Can we regenerate original audio recordings using that latent space?
    \item Can we create new audio synthesis methods using the latent space for sound design applications and composition practices?
    \item Can a DL-based model be used to provide professional grade sound synthesis tools?
    \item How can an audio synthesis architecture using a Deep Learning model provide the user the flexibility to generate sounds of any duration? 
\end{itemize}

 We are also inspired by the definition of music as ``nothing but organized sound''~\cite{varese_liberation_1966} involving sound objects~\cite{schaeffer_traite_1964} that situate on multiple layers~\cite{stockhausen_four_1972}, where any sound can be used to produce music~\cite{luigi_art_1967,varese_liberation_1966}, and strong connections exist between pitch, noise, timbre, and rhythm~\cite{stockhausen_four_1972,smalley_spectromorphology:_1997,roads_microsound_2004,roads_composing_2015}. In that sense, the Latent Timbre Synthesis project builds on our previous work titled Musical Agents based on Self-Organizing Maps (MASOM) \cite{tatar_masom:_2017,tatar_audio-based_2019}. MASOM combines organizing sound samples in latent audio space with statistical sequence models for musical structure. The latent audio space in MASOM is generated by a Self-Organizing Map that organizes a set of audio excerpts. In LTS, we move further by aiming for an audio synthesis framework where we can synthesize sounds that do not exist in the training set.  

Following the research questions and directions above, our contributions presented in this paper include two Variational Auto-encoders (VAEs) to generate a latent space of audio frames. Unlike other Deep Learning architectures such as Generative Adversarial Networks (GANs), VAEs are beneficial for our applications because these architectures can encode an existing audio frame to a latent space, as well as synthesize audio frames from latent vectors. VAEs also allow audio synthesis through interpolation and extrapolation of timbres, by using the latent vectors of audio frames. 

Latent Timbre Synthesis differs from the previous works such as Granma MagNet\cite{akten_grannma_2018} because we prioritize the flexibility to generate audio with any duration, in comparison to outputting audio excerpts of fixed-duration. We think that the flexibility of changing the duration of the generated audio is crucial for our applications, which stands out as another contribution of LTS. Our approach focus on creating a latent space of audio frames, where we can represent an audio recording with any length as a time-series sequence of latent vectors.

The LTS framework consists of three main modules, calculation of wavelet transform based spectrogram representation, latent audio frame space generation using two specific Variational Auto-encoders and inverse synthesis using wavelet-based magnitude spectrogram generated by the decoder of the VAE. We compare the advantages and drawback of two VAE architectures for designing a synthesis tool for composition practices and sound design applications. In addition, we present and share a fully working application, called \textit{interpolate\_two} with a Graphical User Interface (GUI) that allows composers to synthesize audio using timbre interpolation and extrapolation with multiple sounds. In comparison to high computational complexity of previous works mentioned in Section~\ref{sec:background}, the low computational complexity of \textit{interpolate\_two} allows the incorporation of the sound design tool within composition practices and real-time applications. The documentation of the setup of \textit{interpolate\_two} is detailed to guide practitioners of all backgrounds. Our implementation is open-source\footnote{The source code is available at \url{https://www.gitlab.com/ktatar/latent-timbre-synthesis}.}, and sound examples are available\footnote{We provide sound examples at \url{https://kivanctatar.com/Latent-Timbre-Synthesis}.}. We encourage our readers to dive into the code and experiment with the framework for further audio synthesis possibilities.

\section{Related Works}
\label{sec:background}

We limit this section to the previous works that utilize audio spectrogram as an input for the Deep Learning architecture, with the exception of WaveNet. We situate the Latent Timbre Synthesis project within the raw-audio generation applications of Deep Learning, and WaveNet is one of the state of the art systems in the area. We also omit Deep Learning systems for speech synthesis or vocoder applications, such as MelGAN~\cite{kumar_melgan_2019}, while mentioning in Section~\ref{sec:future} how we plan to incorporate them in LTS as a next step in our research.


WaveNet is a Deep Learning architecture that uses an audio corpus for the tasks of music composition, multi-speaker speech and text to speech generation, and speech recognition \cite{oord_wavenet:_2016}. WaveNet applies Convolutional Neural Networks (CNNs) with two strategies to handle temporality of raw audio data: causal convolution and dilation. Causal convolutions ensure that the output only depends on the past observations. Dilated causal convolutions skips a number of inputs on each layer. The number of inputs that are skipped exponentially increases with each layer; hence, the receptive field of the network also increases exponentially \cite{yu_multi-scale_2015}. Note that, the receptive field is the number of neurons that affect a single neuron in deep networks. Oord et al. \cite{oord_wavenet:_2016} tested WaveNet on two audio corpora: the MagnaTagATune dataset and the YouTube piano dataset. The authors point out that ``Even with a receptive field of several seconds, the models did not enforce long-range consistency which resulted in second-to-second variations in genre, instrumentation, volume and sound quality.'' That is, WaveNet struggled to generate long-term variations like in the case of interactive music systems that apply Markov Models \cite[Section~6.1]{tatar_musical_2019}. There has been follow-up research on the WaveNet architecture, where the authors stack multiple WaveNet architecture on top of each other \cite{oord_parallel_2017}, or they combine WaveNet with Vector Quantized Variational Autoencoders \cite{dieleman_challenge_2018}. The main drawback of all WaveNet systems are their computational complexity and high-usage of GPU memory \cite{dieleman_sander_nodate}. The technology requirements of WaveNet compromise its usage in compositional practices, where composers do not necessarily have access to computers with the state of the art GPUs.          

Differentiable Digital Signal Processing (DDSP) is a toolbox made by the Google for researching Digital Signal Processing (DSP) applications of Deep Learning \cite{engel_ddsp_2019}. The authors describe the DDSP Autoencoder, which is a VAE architecture where the input are the Mel-Frequency Cepstral Coefficients (MFCCs) of an audio excerpt. We mention a comparison of using MFCCs and other audio features as the representation of timbre for audio synthesis with VAEs in Section~\ref{sec:spectrograms}. The architecture employs three autoencoders for fundamental frequency (\textit{f-encoder}), loudness (\textit{l-encoder}), and the latent space of timbre (\textit{z-encoder}). The fundamental frequency and the loudness encoders use the CREPE architecture that is originally presented as a pitch detector \cite{hantrakul_fast_2019}. The \textit{z-encoder} architecture is inspired by the ResNet architecture in Computer Vision research \cite{he_deep_2016}. The decoder, on the other hand, controls the input parameters of an additive synthesis module, a subtractive synthesis module, and a reverb. These three synthesis modules generate the final audio. The loss function compares the generated audio with the original one, using a specific function called Multi-Scale Spectrogram Loss, which is similar to comparing the spectrograms of original and generated audio. 

Generative Timbre Spaces project \cite{esling_generative_2018} is perhaps one of the most similar previous study to the Latent Timbre Synthesis project. The application of Generative Timbre Synthesis focuses on generating a latent timbre space of conventional musical instruments. This model uses a VAE where the encoder is a 3-layer feed-forward network with 2000 units in each layer. The latent space has 64 dimensions. The authors introduce a new regularization item in the cost function. The additional regularization loss tries to force the network to satisfy perceptual similarity ratings of conventional musical instruments in Western Classical Music. These perceptual ratings are proposed in previous studies \cite{grey_multidimensional_1977,krumhansl_why_1989,iverson_isolating_1993,mcadams_perceptual_1995,lakatos_common_2000}. The training dataset of Generative Timbre Spaces is audio recordings of conventional musical instruments where each file is an instrument playing a note. The authors takes one frame from each audio file to train the VAE model. Hence, the architecture aims to capture the generalized timbre of a conventional musical instrument instead of the regeneration of an arbitrary audio excerpt. Likewise, the cost function is not suitable to regenerate a dataset with arbitrary audio recordings because there are no perceptual ratings available. We further discuss the issues related to the hyper-parameters of VAE in Generative Timbre Spaces in Section~\ref{sec:deep_learning}.

The DDSP Autoencoder as well as the Generative Timbre Spaces aims for the synthesis applications of conventional music where the model is conditioned to output an audio with a fundamental frequency constraint. In sound design, experimental electronic music, and Sound Art applications, having a fundamental frequency of a sound gesture is rather limiting. The music theory of the contemporary electronic music emphasizes the continuum between noise, pitch, and rhythm \cite{stockhausen_four_1972,luigi_art_1967,smalley_spectromorphology:_1997,roads_microsound_2004,roads_composing_2015}. Hence, in LTS, we aim for a model that could generate any audio so that the composers and practitioners are free to explore the full potential of digital audio synthesis. 



\section{System Design}

\begin{figure}
    \centering
	\includegraphics[width=\columnwidth]{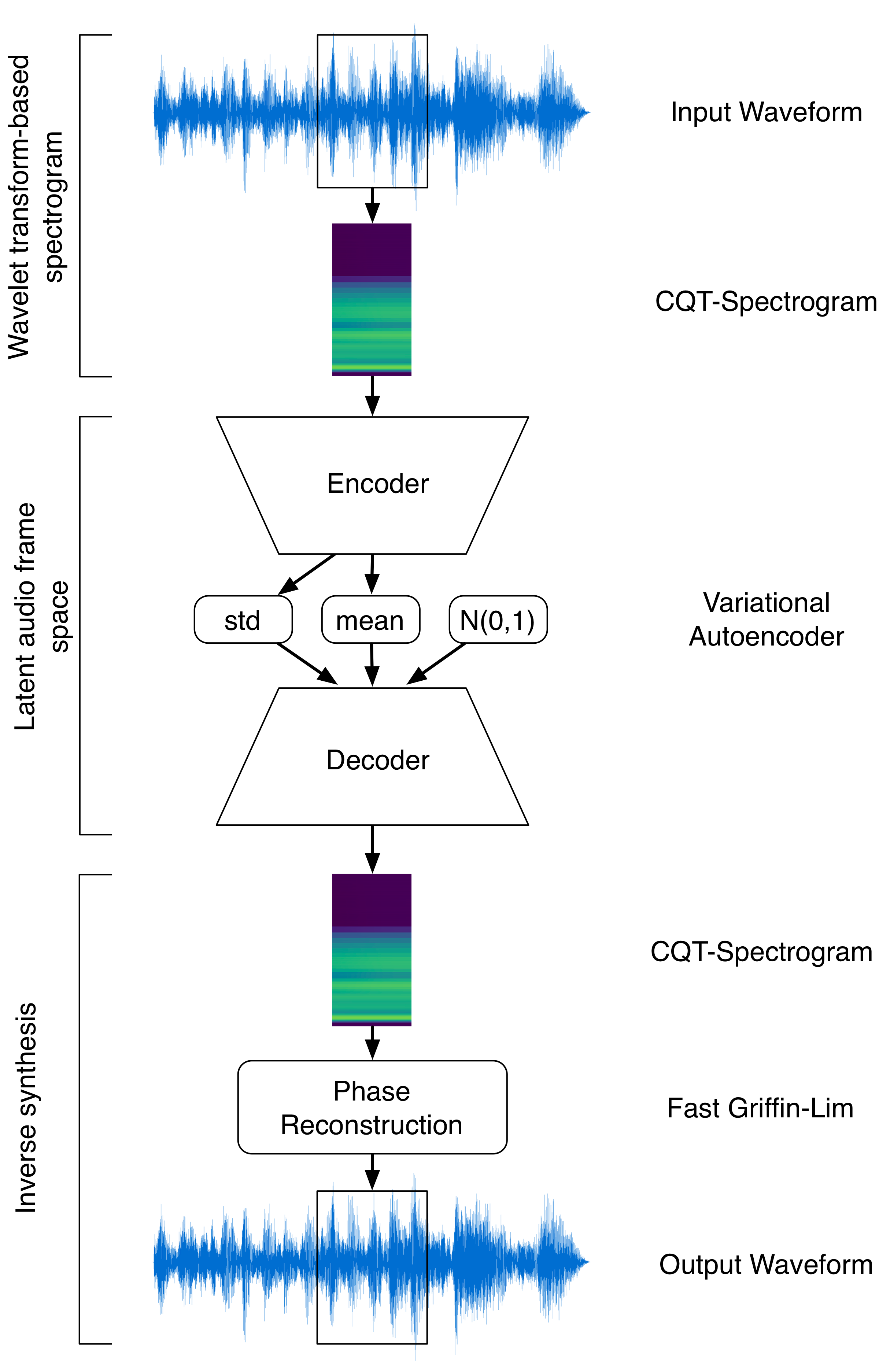}
	\caption{Latent Timbre Synthesis framework}
	\label{fig:timbre_vae}
\end{figure}

The Latent Timbre Synthesis framework consists of three main parts, spectrogram calculation using a type of wavelet transform, audio frame latent space generation using Variational Autoencoders, and inverse synthesis of audio using the magnitude spectrogram generated by the decoder of VAE (Figure~\ref{fig:timbre_vae}). 

\subsection{Wavelet Transform based Spectrograms}
\label{sec:spectrograms}
The audio feature extraction module generates spectrogram frames using a type of wavelet transform for audio, called Constant-Q Transform (CQT) \cite{schorkhuber_constant-q_2010}, which gained popularity in Music Information Retrieval (MIR) research in the recent years. CQT, as well as its variant Non-stationary Gabor Transform (NSGT) \cite{velasco_constructing_2011}, have been compared to the other audio features such as Mel-Frequency Cepstral Coefficients (MFCC); and previous studies showed that CQT and NSGT could perform better in MIR applications such as segmentation and musical structure analysis \cite{nieto_systematic_2016}. Naturally, segmentation and musical structure analysis tasks require computing the audio similarity~\cite{muller_fundamentals_2015}; thus, they are suited to create a latent space of audio frames. 

\begin{table}
    \caption{A previous study provided a comparison of audio frame reconstruction losses with Variational Autoencoders using spectrograms with fixed-length windows and wavelet transform based spectrograms \protect\cite{esling_generative_2018}.}
    \label{tab:spectrogram_comparison}    
    \begin{tabular}{llll}
    \hline\noalign{\smallskip}
         & Spectrogram & $logp(x)$ & $||x-\tilde{x}||^2$ \\
         \noalign{\smallskip}\hline\noalign{\smallskip}
         \multirow{2}{*}{Fixed Window} & STFT & -1.9237 & 0.2412 \\
         & DCT & 4.3415 & 2.2629 \\
         \hline
        \multirow{3}{*}{Wavelet Transform} & CQT & -2.8723 & 0.1610 \\
        & NSGT-MEL & -2.9184 & 0.1602 \\
        & NSGT-ERB & -2.9212 & 0.1511 \\
        \noalign{\smallskip}\hline 
    \end{tabular}
\end{table}

A previous work \cite{esling_generative_2018} compared spectrograms computed with fixed windows and wavelet transform based spectrograms for applications of latent audio frame space generation using Deep Learning (DL). This comparison included Short-Time Fourier Transform, Discrete Cosine Transform, Constant-Q Transform (CQT), and Non-Stationary Gabor Transform (NSGT) variations using different frequency scales. The wavelet based transforms in this previous study were CQT and NSGT variants. The authors \cite{esling_generative_2018} found that wavelet transform based spectrogram representations perform better than the spectrograms calculated using fixed-length windows in regards to the log-likelihood and mean quality of the audio frame reconstructions, as shown in Table~\ref{tab:spectrogram_comparison}; while the audio frame reconstructions of wavelet transform based spectrograms gave similar results. We utilize CQT in comparison to other wavelet-based spectrograms in Table~\ref{tab:spectrogram_comparison} because a python library for audio analysis, titled Librosa\footnote{\url{https://librosa.github.io/librosa/}} \cite{mcfee_librosa_2015}, includes a CQT and inverse CQT implementation \cite{schorkhuber_constant-q_2010} combined with a Fast-Griffin-Lim phase estimation \cite{perraudin_fast_2013} that we explain in Section~\ref{sec:inverse_synthesis}. We aimed that the LTS framework would be available for composers and sound designers of all technical levels. Hence, we prioritized the options that simplify the installation of the LTS framework.  


We delve into the details of Constant-Q Transform in the following. We can calculate the CQT of an audio recording \cite{schorkhuber_constant-q_2010}, a discrete time domain signal $x(n)$, using the following formula:

\begin{equation}
    X^{CQ} (k,n) = \sum_{j=n- \lfloor N_k /2 \rfloor} ^{n+ \lfloor N_k /2 \rfloor } x(j) a_k ^*(j-n+N_k /2)
\end{equation} 
where $k$ represents the CQT frequency bins with a range of $[1, K]$, and $X^{CQ} (k,n)$ is the CQT transform. $N_k$ is the window length of a CQT bin, that is inversely proportional to $f_k$ that we define in equation~\ref{eq:fk} Notice that, $ \lfloor \cdot \rfloor$ is the rounding towards negative infinity. $a_k ^ *$ is the negative conjugate of the basis function $a_k (n)$ and, 

\begin{equation}
    a_k (n) = \frac {1}{N_k} w (\frac{n}{N_k}) exp [-i 2\pi n \frac{f_k}{f_s} ] 
\end{equation}
where $w(t)$ is the window function, $f_k$ is the center frequency of bin $k$, and $f_s$ is the sampling rate. CQT requires a fundamental frequency parameter $f_1$, which is the center frequency of the lowest bin. The center frequencies of remaining bins are calculated using, 

\begin{equation}
\label{eq:fk}
    f_k = f_1 2 ^ {\frac{k-1}{B}}
\end{equation}
where $B$ is the number of bins per octave.  

\begin{figure}
\begin{subfigure}{\columnwidth}
    \centering
	\includegraphics[width=\columnwidth]{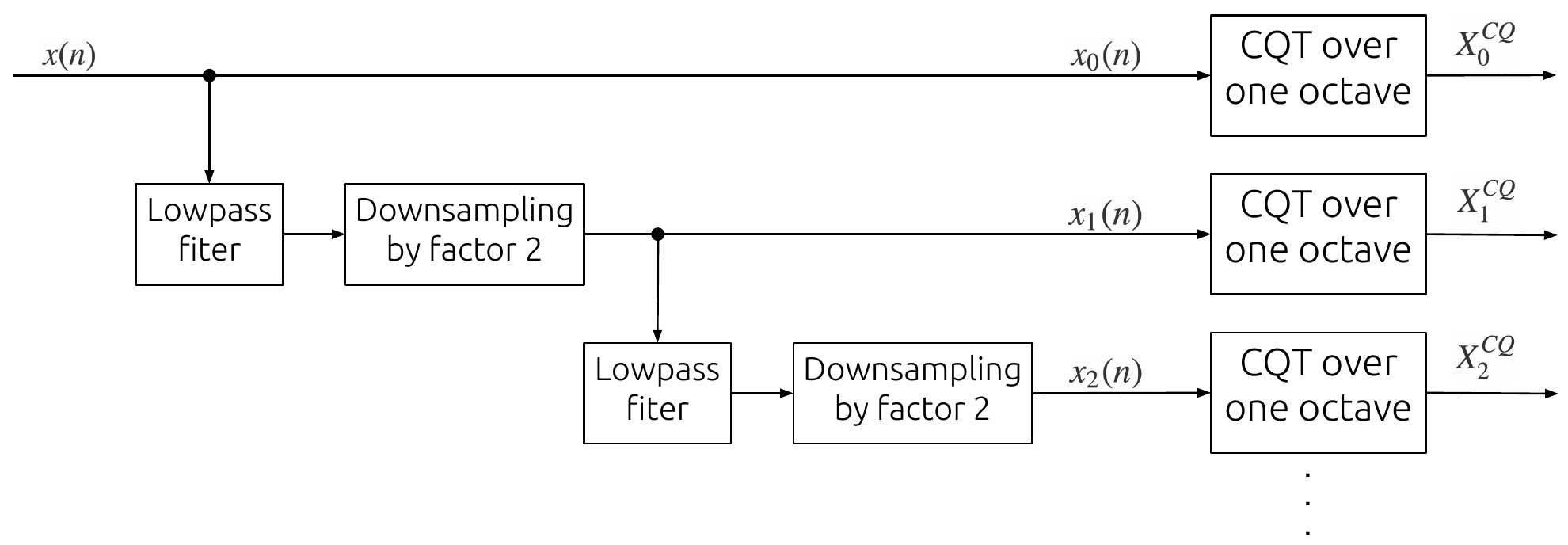}
	\caption{Constant-Q Transform}
	\label{fig:cqt}
\end{subfigure}

\begin{subfigure}{\columnwidth}
    \centering
	\includegraphics[width=\columnwidth]{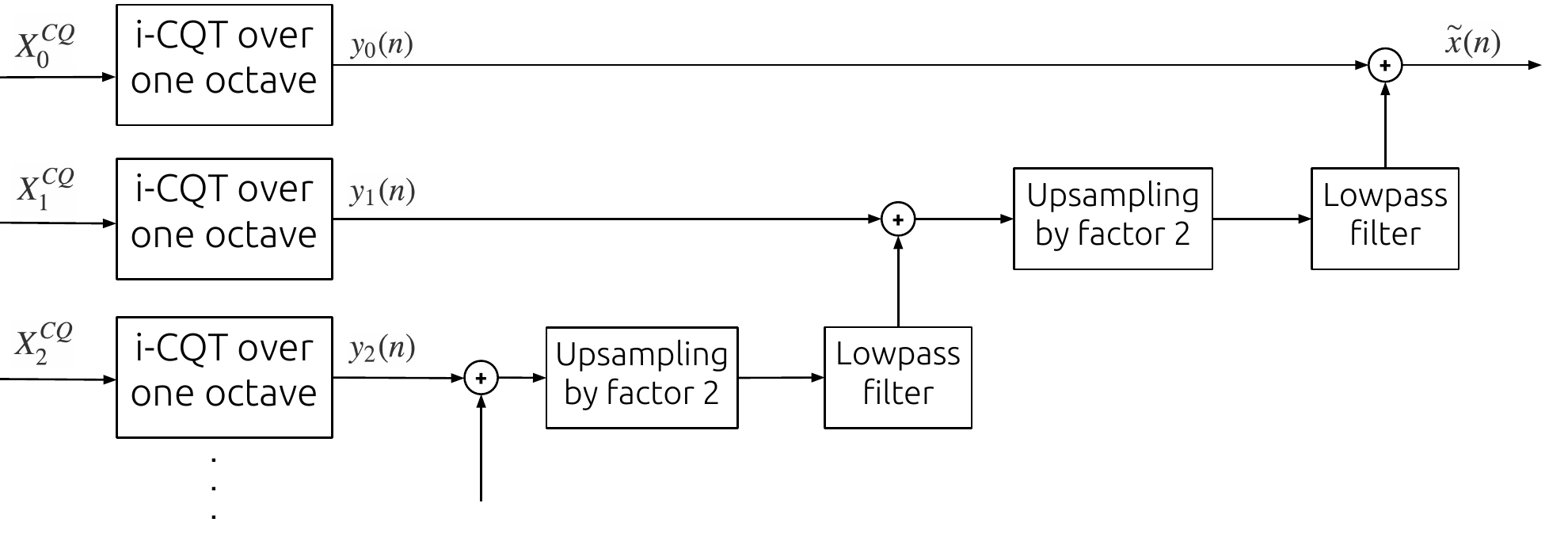}
	\caption{Inverse Constant-Q Transform}
	\label{fig:icqt}
\end{subfigure}
\caption{A fast algorithm to compute CQT and i-CQT, described in \cite{schorkhuber_constant-q_2010} and implemented in Librosa \cite{mcfee_librosa_2015}.}
\label{fig:cqt_algorithm}
\end{figure}

CQT is a wavelet-based transform because the window size is inversely proportional to the $f_k$ while ensuring the same Q-factor for all bins $k$. We can calculate the Q-factor using,

\begin{equation}
\label{eq:Q}
    Q = \frac{qf_s}{f_k (2^\frac{1}{B} - 1)}
\end{equation}
where $q$ is scaling factor with the range [0,1] and equals to $1$ as the default setting. We direct our readers to the original publication for the specific details of the CQT \cite{schorkhuber_constant-q_2010}, which also proposed a fast algorithm to compute CQT and inverse CQT (i-CQT), given in Figure~\ref{fig:cqt_algorithm}.    

All experiments in this paper used the same audio feature extraction configurations. We tried several parameters to find the configuration that could give the least amount of audio artifacts with the pipeline of calculating the CQT spectrogram and then reconstructing the audio back using the magnitude spectrogram of CQT combined with the phase estimation algorithm. Notice that these artifacts would appear even with an ideal DL model because the inverse audio synthesis introduces these artifacts to the LTS. We used 16-bit and 44.1 kHz stereo or mono audio recordings. We converted the stereo files to mono first, and then calculated the CQT spectrograms using a hop-size of 128 samples. The $f_1$ parameter was 32.7 Hz that corresponds to the musical note, C1. $q$ value in equation \ref{eq:Q} was equal to $1$, and the window function was ``hann''\footnote{\url{https://docs.scipy.org/doc/scipy/reference/generated/scipy.signal.windows.hann.html}}. CQT included 48 bins per octave for a total range of 8 octaves; which sums up to a 384 bins in total. Hence, the input of the DL models are vectors with 384 dimensions. These parameters resulted in the least amount of artifacts in our experiments. Our source code is flexible to change these parameters, and we encourage our readers to try different parameter configurations.

\subsection{Deep Learning in Latent Timbre Synthesis}
\label{sec:deep_learning}

\subsubsection{Autoencoders and Variational Autoencoders}

\begin{figure}
    \centering
	\includegraphics[width=\columnwidth]{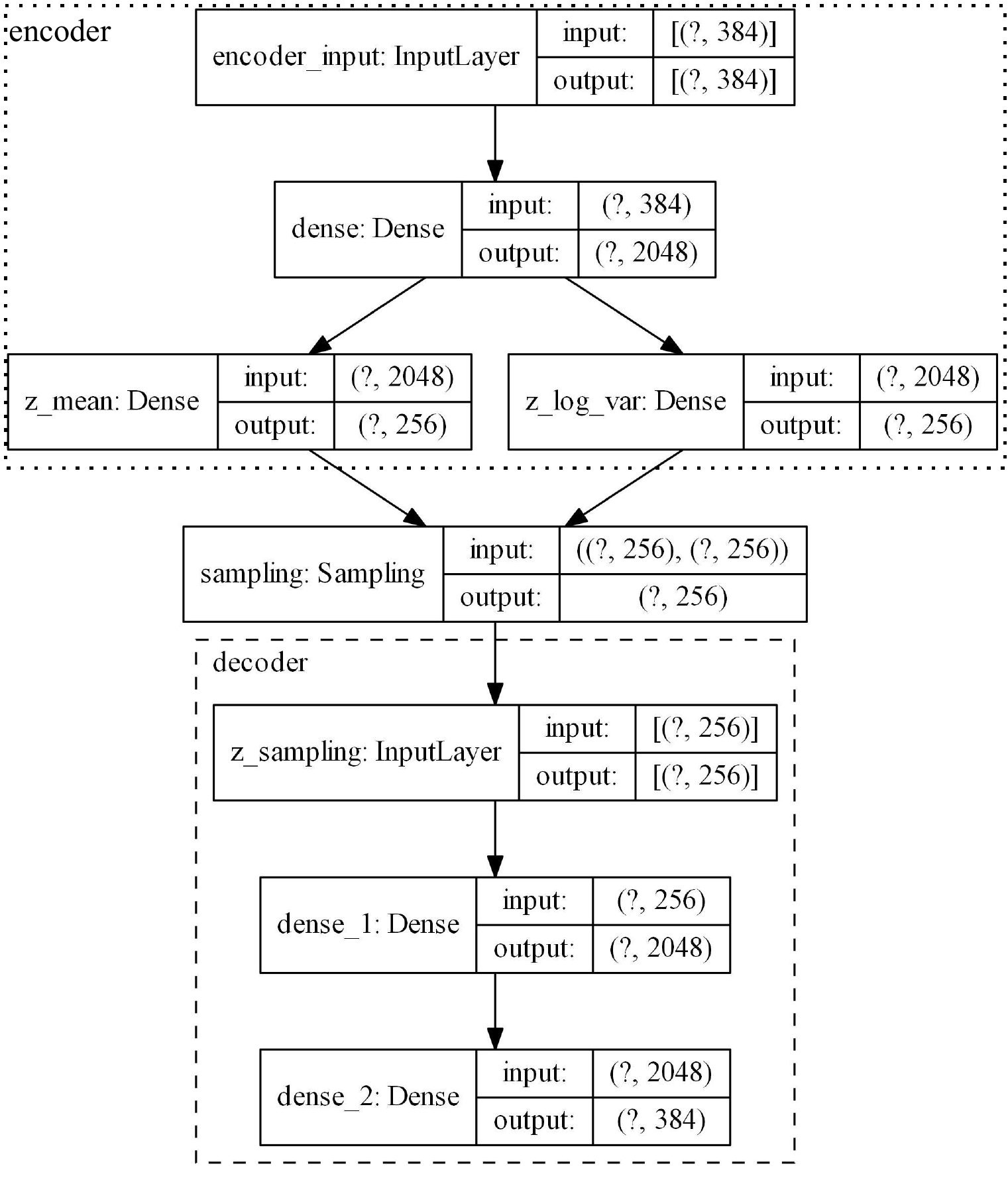}
	\caption{The first Deep Learning architecture available in Latent Timbre Synthesis}
	\label{fig:vae_model}
\end{figure}

Autoencoders are Deep Learning architectures for generative modelling. The architecture consists of two main modules: an encoder and a decoder (Figure \ref{fig:timbre_vae} and \ref{fig:vae_model}). The encoder maps the input data $x \in R^L$ to a latent vector $z \in R^M$ where $z=encoder(x)$, and $M<L$. The decoder aims to convert a latent vector back to the original data, and ideally, $decoder(encoder(x)) = x$.  The vanilla Autoencoder architecture encodes the input data vector to a single point, that is the latent vector. In comparison, Variational Autoencoder (VAE) is an improved version of the Autoencoder architecture that converts the input data vector to a stochastic distribution over the latent space. This difference is also referred as the ``reparametrization trick'' \cite{kingma_auto-encoding_2014,kingma_introduction_2019,sonderby_how_2016}. 

In VAE, the encoder tries to generate a latent space by approximating $p(z|x)$ while the decoder tries to capture the true posterior $p(x|z)$. The vanilla VAE approximates $p(z|x)$ using $q(z|x) \in Q$ with the assumption that $p(z|x)$ is in the form of a Gaussian distribution $N(0,I)$. This approximation is referred in the literature as \textit{Variational Inference} \cite{kingma_auto-encoding_2014}. Specifically, the encoder outputs the mean $\mu _M$ and the co-variance $\sigma _ M$ as the inputs of the Gaussian distribution function $N(z; \mu_M, \sigma^2 _M I)$ over a latent space with $M$ number of dimensions. Hence, the encoder approximates $p(z|x)$ using $q^* (z|x) = N(z; f(x),g(x) ^2 I)$ where $\mu _M = f(x)$, $f \in F$, $\sigma _ M = g(x)$, and $g \in G$. The decoder's input, the latent vector $z$ is sampled from the latent distribution $q(z) = N(z; f(x),g(x) ^2 I)$. Hence, the loss function consists of the reconstruction loss and the regularization term of Kullback-Leibler divergence between $q^* (z|x)$ and $p^* (z)$,

\begin{equation}
\label{eq:vae_loss}
    L_{f,g} = \mathbb{E} _{q^* (z)}[logp^* (x|z)] - \alpha \cdot D_{KL}[q^* (z|x)||p^* (z)]
\end{equation}

We direct our readers to the original VAE publication \cite{kingma_auto-encoding_2014} for the mathematical induction of the loss function in equation \ref{eq:vae_loss} \cite{kingma_introduction_2019}. Note that, some previous works introduced additional regularization terms to the loss function to condition the VAE further, such as the introduction of perceptual ratings of musical instruments in \cite{esling_generative_2018}. 

The LTS framework focuses on Variational Autoencoders in comparison to Generative Adversarial Networks because we aim for audio synthesis by interpolation and extrapolation in the latent space of audio frames (see Section \ref{sec:interpolate_two}). The input vectors of the VAE model are CQT vectors calculated from one audio frame where the window size varies with the frequency bins. We aim to generate a latent space of audio frames so that we can synthesize audio with any duration. Previous systems such as Grannma MagNet \cite{akten_grannma_2018} utilizes audio excerpts with fixed-duration, where the training input vectors of deep learning model are 2D audio spectrograms with time along the x-axis and frequency along the y-axis. This design choice constraints these DL models to limited applications such as generating a fixed-length audio excerpt. Our approach differs from the previous systems because the training observation of DL model in LTS is one audio-spectrum vector that is calculated from one audio-frame. This allows LTS to generate audio with any duration in sound design applications. 

We focus on two Deep Learning (DL) architectures in the current version of the Latent Timbre Synthesis framework. Both models are Variational Autoencoders (VAE); however, the layers and model parameters differ. In the first VAE model, we were inspired by the previous work \cite{esling_generative_2018} where the authors trained a Variational Autoencoder to generate conventional musical instrument timbres with digital audio synthesis. 

We initially tried the VAE architecture with the same hyper-parameter settings that were used in the Generative Timbre Spaces project. The setting of Generative Timbre Spaces \cite{esling_generative_2018} were unsuccessful in our experiments. The model could not learn to regenerate the audio recordings in the training dataset, and could only generate noise. Upon further investigation, we found that the Kullback-Leibler Divergence regularization term multiplier caused the issue. In Generative Timbre Spaces, the authors increase the multiplier from $0$ to $2$ during the first $100$ epochs of the training, following the warm-up procedure \cite{sonderby_how_2016}. We suspect that the KLD multiplier range of $[0,2]$ is specific to the application of Generative Timbre Spaces where the training dataset consists of audio files with distinct harmonic content and low noisiness in the spectrum. Furthermore, the training dataset size of Generative Timbre Spaces is rather small, less than 100 MB, whereas we work with GBs of audio to train the LTS models. For example, the \textit{erokia} dataset that we provide with our source code includes 2 GBs of audio that corresponds to $3,084,591$ audio frames as training data for LTS models, using a hop-size of $128$ samples for the CQT calculation. 

We found that the range of $[0,2]$ for KLD multiplier was too high for our application and prohibited the VAE to learn. In addition, the warm-up procedure had adverse effects on the learning. We further investigated this issue using the MNIST dataset \cite{lecun_mnist_nodate}. MNIST gave us a visual understanding of the effect of KLD multiplier. The images in MNIST are $28*28$ pixels, adding up to a total of $784$ pixels. This is similar to LTS where the input of the VAE is a vector of $384$ dimensions. 

Like in the case of our tests with audio spectrograms, we obtained similar results with the MNIST data. Higher KL-divergence values as well as the warm-up procedure significantly deteriorated the reconstructions of the trained model.  

\begin{figure*}
\begin{subfigure}{\columnwidth}
    \centering
	\includegraphics[width=\columnwidth]{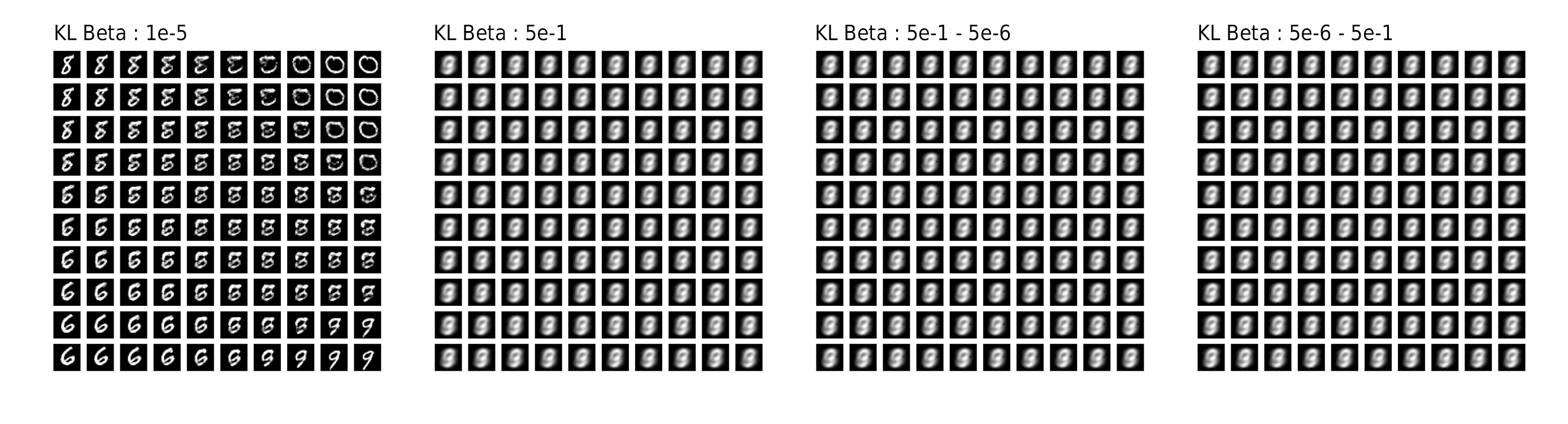}
	\caption{The effect of the KLD multiplier on the reconstructions}
	\label{fig:MNIST:interpolations}
\end{subfigure}

\begin{subfigure}{\columnwidth}
    \centering
	\includegraphics[width=\columnwidth]{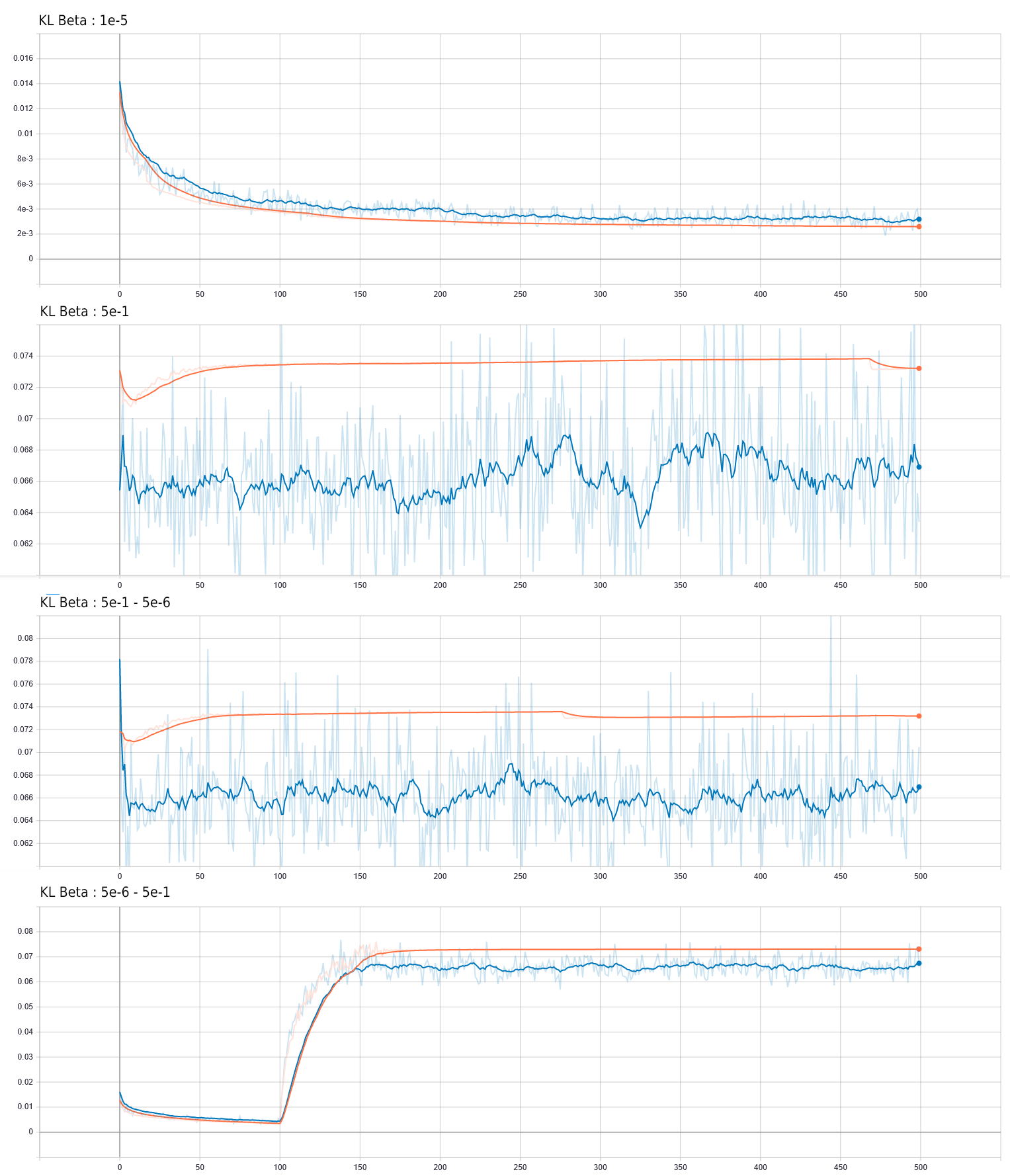}
	\caption{The loss values during training where x-axis is the training epoch and y-axis is the loss value}
	\label{fig:MNIST:logs}
\end{subfigure}
\caption{The effect of the KLD multiplier in the loss function on Variational Autoencoder training}
\label{fig:MNIST}
\end{figure*}

Figure \ref{fig:MNIST:interpolations} and \ref{fig:MNIST:logs} show the effect of KLD multiplier on the training of the Variational Autoencoder. We used the same architecture depicted in Figure \ref{fig:timbre_vae} and only changed the input and output dimensions to $784$ that corresponds the flattened vector of $28x28$ images in the MNIST dataset. In addition, we tested the warm-up procedure and the reverse settings of the warm-up procedure; however, both were detrimental to the training in our experiments. We proceeded our experiments with KLD values around $1e-5$, given the success of this setting with the MNIST dataset. 

All VAE architectures in LTS use decoder networks that are the reversed replicas of the encoder networks, as in most cases of VAEs. The first VAE model in LTS is a feed-forward network with two Dense layers including 2048 neurons, The \textit{dense, dense\_1,} and \textit{dense\_2} layers in Figure \ref{fig:vae_model} apply Rectified Linear Units (ReLU)  as the neuron activation functions.

\begin{figure}
    \centering
	\includegraphics[width=0.65\columnwidth]{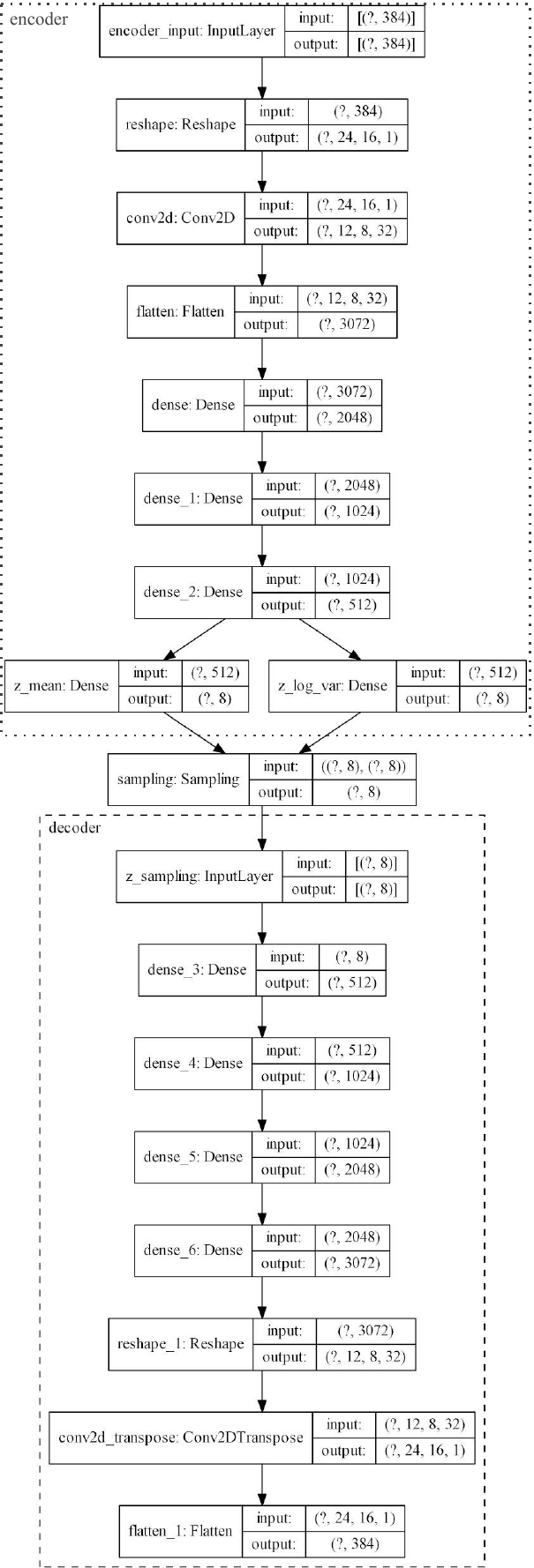}
	\caption{The second Variational Autoencoder architecture available in Latent Timbre Synthesis}
	\label{fig:cvae_model}
\end{figure}

We train the network for 2000 epochs, while the improvements after epoch 50 are rather in the exploitation\footnote{Exploration and exploitation are two search strategies in optimization applications \cite[Section 5.3]{tatar_automatic_2016}.} phase of the learning, and help to minimize the floor noise in the generated CQT magnitude spectrogram. The learning rate is $1e^{-4}$ and the KLD multiplier in the cost function is $5e^{-4}$. The learning rate and the KLD multiplier parameters are dependent on the training dataset. We recommend the readers who would be interested to try their own dataset to start with the hyper-parameter settings above, and change the parameters when needed. 


The latent space of our first VAE model consists of $256$ dimensions. This number is still acceptable given that the previous work \cite{esling_generative_2018} used a $64$ dimensional latent space to cluster a much smaller range of conventional musical instrument timbres. Yet, we explored the possibility to find a deeper network that could generate a latent space with a smaller number of dimensions. 

The second VAE network that we present in this paper aimed to decrease the number of latent space dimensions of the first network using a deeper architecture, shown in Figure \ref{fig:cvae_model}. We pursued an iterative design procedure where we tried to decrease the number of dimensions of the latent space while maintaining the reconstruction quality and trying to achieve lower final loss values in the training runs. Our experiments aimed for a 8 dimensional latent space, and we first tested increasing the number of dense layers. Our experiments found an optimum of 4 dense layers where the number of neurons is changing by two-folds each layer (Figure \ref{fig:cvae_model}). In addition to the Dense layers, we also tried adding convolutional layers on top of the Dense layers. We imagined that the convolutional layers could grasp the relationships between frequency bins of a given CQT vector, such as harmonics. The convolutional layer generates a 2-dimensional vector with size 24x16, which is in relation to 48 bins per octave for 8 octaves while trying to remain as close as to a square. Notice that, each row in the 2D vector generated by the convolutional layer corresponds to half of an octave. We imagined that this could further help the network to capture the harmonic content in the CQT vectors. Our experiments found an optimum of one additional convolutional layer with $32$ filters and a kernel size of $3$ and a stride size of $2$ on top of the dense layers, shown in Figure \ref{fig:cvae_model}. The final second network could decrease the number of latent space dimensions while giving a low final loss value. However, the network introduced a consistent floor noise sound in the reconstructions. The introduction of normalization techniques such as batch, instance, or layer normalization could not eliminate the floor noise in the reconstructions. Our findings suggest a balance between the number of latent space dimensions and the floor noise in the reconstructions where increasing the number of latent space dimensions is the solution to eliminate the noise. 

Given that audio quality is of great importance in composition tasks, we focused on using and disseminating the first network given in Figure \ref{fig:vae_model}. Additionally, the low computational complexity of the first network can be an advantage when we combine the VAE with Deep Learning models for time-series sequence generation in our future work, detailed in Section \ref{sec:future}.  

\subsection{Inverse Synthesis and Audio Reconstruction}
\label{sec:inverse_synthesis}

This version of LTS uses the Fast Griffin-Lim \cite{perraudin_fast_2013} phase estimation algorithm (GLA) for generating audio from CQT magnitude spectrograms that the VAE decoder outputs.  Briefly, the GLA estimates the phase component of a magnitude spectrogram by iterating the inverse synthesis and the spectrogram calculation multiple times, initially proposed in \cite{griffin_signal_1984} and shown in Algorithm \ref{alg:GLA} \cite{perraudin_fast_2013}. Given an an audio signal $x(n)$ and its frequency transform $X(i)$,

\begin{algorithm}
    \caption{Griffin-Lim Algorithm}
	\begin{algorithmic}[1]
        \State \textbf{Set:} $\angle X_0(i)$ 
        \State \textbf{Initialize:} $X_0(i) = |X(i)| \cdot e^{j\angle X_0(i)} $
        \For {$n=1,2,\ldots, N$}
            \State $X_n(i) = T(IT(|X(i)| \cdot e^{j\angle X_{n-1}(i)}))$
        \EndFor
        \State $\hat x(n) = IT(X_N(i)) $
    \end{algorithmic}
    \label{alg:GLA}
\end{algorithm} where $N$ is the total number of GLA iterations, $T$ and $IT$ is the frequency transform and inverse frequency transform function respectively; such as Short-Fourier Transform, or Constant-Q Transform in our case. Note that, the space of audio spectrograms is a subset of the complex number space. The iterative process of Griffin-Lim moves the complex spectrogram of the estimated signal $\hat x(n)$ towards the complex number space of audio signals in each iteration, as proven in \cite{griffin_signal_1984}. 




The Fast Griffin-Lim algorithm (F-GLA) \cite{perraudin_fast_2013} is a revision of the original Griffin-Lim algorithm. 

\begin{algorithm}
    \caption{Fast Griffin-Lim Algorithm}
	\begin{algorithmic}[1]
        \State \textbf{Set:} $\angle X_0(i)$ 
        \State \textbf{Initialize:} $X_0(i) = |X(i)| \cdot e^{j\angle X_0(i)} $
        \State \textbf{Initialize:} $Y_0(i) = T(IT(|X(i)| \cdot e^{j\angle X_0(i)})) $
        \For {$n=1,2,\ldots, N$}
            \State $Y_n(i) = T(IT(|X(i)| \cdot e^{j\angle X_{n-1}(i)}))$
            \State $X_n(i) = Y_n(i) + \alpha (Y_n(i) - Y_{n-1}(i)) $
        \EndFor
        \State $\hat x(n) = IT(X_N(i)) $
    \end{algorithmic}
    \label{alg:FGLA}
\end{algorithm}
where $\alpha$ is a constant. A previous study \cite{perraudin_fast_2013} showed that the F-GLA revision significantly improves signal-to-noise ratio (SNR) compared to the GLA, where the setting $\alpha = 1$ resulted in the highest SNR value. 

The F-GLA module has the highest computation time in the LTS framework. We are aware that a revision of this module using another Deep Learning architecture for vocoder applications can improve the computational complexity of LTS while making LTS more lightweight within real-time applications. We further address this in the Section \ref{sec:future}.

This concludes the explanation of the LTS architecture, where we covered our specific choice of Constant-Q Transform, two Variational Autoencoder architectures, and the inverse synthesis using the GLA phase estimation. In the following, we introduce the \textit{interpolate\_two} algorithm. 

\section{Interpolation and Extrapolation in Latent Timbre Space}
\label{sec:interpolate_two}

The first sound design application of Latent Timbre Synthesis is the \textit{interpolate\_two} framework that allows composers to synthesize sounds using interpolation and extrapolation with two sounds. The framework requires a trained model to generate sounds using the latent audio frame space. The user can select the duration of the generated sound, and can choose an excerpt from two audio files. These two excerpts have the same duration. The algorithm uses these two excerpts for synthesis with interpolation and extrapolation in the timbre space. The interpolation amount sets how much of the latent vector is copied from one of the audio excerpts. For example, 30\% interpolation is adding 30\% of the latent vectors of the first audio and 70\% of the latent vectors of the second audio. The percentages above 100 or below 0 corresponds to extrapolations. For instance, 120\% is moving 20\% away from the second audio in the direction the latent vector that points from audio 1 to audio 2.
The algorithm synthesizes the audio by calculating every audio frame using inverse synthesis from a generated spectrogram. Hence, the user sets interpolation amounts for each latent vector that corresponds to one audio frame. The user can draw an interpolation curve to change the interpolation percentage in time using the LTS framework.
\begin{figure}
	\includegraphics[width=\columnwidth]{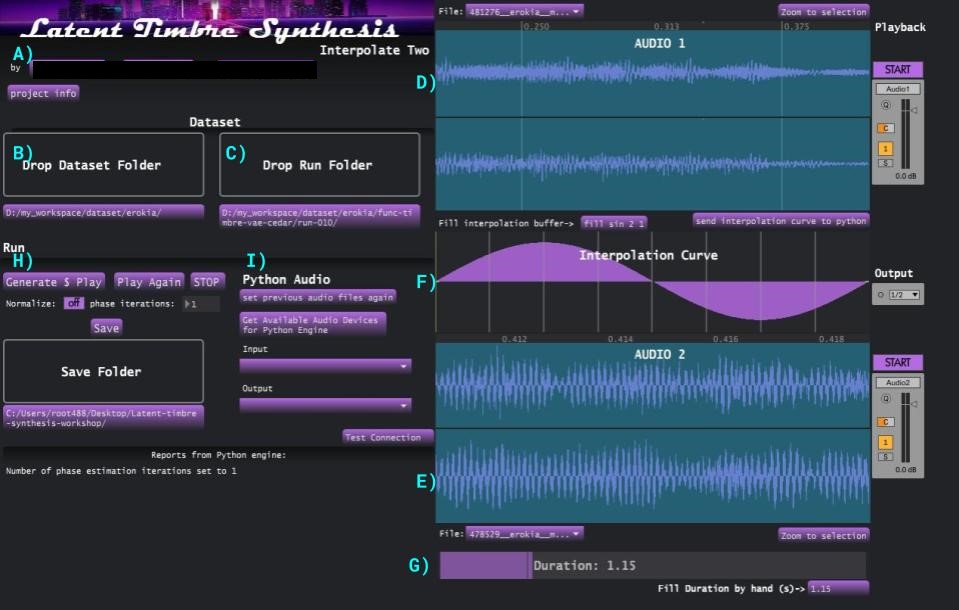}
	\caption{The Max GUI of the \textit{interpolate\_two} application}
	\label{fig:interpolate_two_gui}
\end{figure}

The application, \textit{interpolate\_two} consists of two components: Max GUI and the python engine. The Max GUI handles user interactions while the python engine reacts to OSC messages coming from Max. The python engine runs the deep learning model and audio feature extraction (CQT calculations), as well as inverse synthesis that generates audio from CQT magnitude spectrogram combined with Fast Griffin-Lim phase estimation. 

The framework of \textit{interpolate\_two} is compatible with all variations of the VAE architectures given in Figure \ref{fig:vae_model} and \ref{fig:cvae_model}. The B an C regions in the GUI in Figure \ref{fig:interpolate_two_gui} load the datasets and models that are produced within a particular run with the dataset. Using the regions D and E, the user can select two audio files from a dataset to choose excerpts. The file dropdown menu allows to select an audio file from the dataset. "Zoom to selection" sets the view to the selection area. Clicking \textit{ctrl} (or \textit{cmd} on macos) and then dragging the mouse up \& down on the waveform views applies zoom in \& out. The waveform in the region F sets apply interpolation (or extrapolation) amounts per frame, where x-axis is the time and y-axis sets the interpolation percentage for a frame. When both waveform views are zoomed to the selection, the x-axis of the interpolation curve corresponds to the x-axis of the waveform. The interpolation curve view is a [waveform\textasciitilde ] object. The "Vertical Zoom" parameter in the inspector of this Max object sets the maximum interpolation/extrapolation amount. The default maximum is 1.3; hence, [1.0,1.3] and [-1.0, -1.3] are the extrapolation regions. It is possible to extrapolate even more by changing the vertical zoom parameter; however, the higher amounts are likely to give audible distortions. The normalize toggle in region H allows the user to normalize the generated audio to prevent audio distortions or extremely high audio volumes while exploring extrapolation possibilities. 

The section H send messages to the python engine to handle output generation. "Generate \& Play" initiates the python engine to synthesize a sound using the current interpolation curve and the audio selections. "Play Again" plays the previous generated sound, without going through the deep learning calculation. "STOP" immediately stops the audio coming out of the python engine. Phase iterations sets the number of iterations of the Fast Griffin-Lim algorithm. Higher number of iterations (max. 64) gives better results; however, the calculation takes significantly longer. The phase estimation algorithm is the bottleneck of computational complexity of this framework. Still, the calculation of the audio takes 50\% of the audio duration with phase iteration set to 1. That is, calculating a 2-second sound takes around 1-second on a laptop with NVIDIA RTX 2080 Max-Q GPU and 2.20 GHz Intel i7-8750H CPU.


\section{Future Work}
\label{sec:future}

We are currently finalizing our VAE model analysis where we visualize the latent audio frame space. In addition, we are conducting a qualitative study where composers and practitioners used the LTS framework for sound design applications. We aim to obtain a better understanding of the creative potential of these algorithms by evaluating them in real-world application scenarios within composition practices. Our future work includes a publication where we plan to include the DL model analysis and the qualitative study, as well as a compilation album release.  

\section*{Acknowledgements}
This research has been supported by the \href{http://www.snf.ch/en/Pages/default.aspx}{Swiss National Science Foundation}, \href{https://www.nserc-crsng.gc.ca/NSERC-CRSNG/Council-Conseil/index_eng.asp}{Natural Sciences and Engineering Research Council of Canada}, \href{https://www.sshrc-crsh.gc.ca/home-accueil-eng.aspx}{Social Sciences and Humanities Research Council of Canada}, and \href{https://www.computecanada.ca/home/}{Compute Canada}.

\bibliographystyle{spmpsci}      
\bibliography{proposal}   

\end{document}